\documentclass[12pt]{article}
\usepackage{amsmath,latexsym}
\usepackage{graphicx}
 \def\cen{\centerline}

\begin{document}

\setlength{\unitlength}{1mm}

\title{Wormholes with varying equation of state parameter }
 \author{\Large$F.Rahaman^*$, $M.Kalam^{\ddag}$ , $M Sarker^*$ and
$S. Chakraborty^\dag$ }

\date{}
 \maketitle
 \begin{abstract}
                  We propose wormholes solutions by assuming space
                  dependent equation of state parameter. Our
                  models show that the existence of wormholes is
                  supported by phantom energy.
                   Here, the phantom energy is characterized by variable
                  equation state parameter.
We show that the averaged null energy condition ( ANEC )
                  violating phantom energy can be reduced as desired.

  \end{abstract}


 \bigskip
 \medskip
  \footnotetext{ Pacs Nos :  04.20 Gz,04.50 + h, 04.20 Jb   

                               $*$Dept.of Mathematics, Jadavpur University, Kolkata-700 032, India\\
                                  E-Mail:farook\_rahaman@yahoo.com

                             $\ddag$Dept. of Phys. , Netaji Nagar College for Women ,
                                          Regent Estate,
                                          Kolkata-700092, India

  $\dag$Dept. of Maths. , Meghnad Saha Institute of Technology,
                                           Kolkata-700150, India
                              }

    \mbox{} \hspace{.2in}

 \cen{ \bf 1. INTRODUCTION }

A Wormhole is a 'tunnel' through curved spacetime, connecting two
widely separated regions of our Universe or even of different
Universe. In a pioneer work, Morris and Thorne [1] observed, to
hold a wormhole open, one has usually used an exotic matter, which
violates the well known energy conditions. The exotic matter is a
hypothetical form of matter that violates the weak or null energy
conditions. In last few years, exotic matter has been becoming an
active area of research in wormhole physics [2]. Since all known
matters obey the null energy condition, $ T_{\mu\nu} k^\mu k^\nu >
0 $, where $T_{\mu\nu}$ is the energy stress tensor and $k^\mu$
any null vector, several authors [3] have considered scalar tensor
theories to build wormhole like spacetime with the presence of
ordinary matter in which scalar field may play the role of exotic
matter. In an interesting paper, Vollick has shown how to produce
exotic matter using scalar field [4].  Recent astrophysical
observations indicated that the Universe at present is
accelerating . There are different ways of evading these
unexpected behavior . Most of these attempts focus on Alternative
gravity theories or the supposition of existence of a hypothetical
dark energy with a positive energy density and a negative pressure
[5].

The matter with the property, energy density, $\rho > 0 $ but
pressure $ p < - \rho < 0 $ is known as Phantom Energy. The idea
of phantom was proposed by Caldwell [6] to describe acceleration
state of the Universe.  As  phantom energy violates the null
energy condition what is needed to support traversable wormholes.
So phantom energy may play a possible role for constructing
wormhole like spacetime.

Several authors have recently discussed the physical properties
and characteristics of traversable wormholes by taking Phantom
Energy as source [7]. Recent observational analysis involving
X-ray luminosity of galaxy clusters and SNe type Ia data suggest
that we live in a flat Universe and its present acceleration stage
driven by a dark energy component whose equation  of state may
evolve in time [8]. Several authors have studied cosmological
models assuming variable equation of state parameter[9]. Since in
the literature of wormhole physics, this dark energy component is
known as phantom energy, in this article, we propose wormhole
solutions supported by phantom energy where equation of state
parameter is a function of radial coordinate rather than a
constant. The present work falls into two categories. In the first
one, we provide phantom energy matter sources that produce
wormhole like geometry. In this category, we discuss two toy
models. In the second one, we are trying to search phantom energy
matter sources that produce some specified wormhole like
structures. In this category, we provide  two specific toy models
of wormholes. In all cases, we have established a matching of each
interior wormhole metric with an exterior Schwarzschild metric.

The layout of the paper as follows : In the second section, we
shall present the model of our system. In section three, we shall
provide four toy models of the wormholes. Section four is devoted
to a brief summary and discussion.
\\
\\
\\
 \cen{ \bf 2. \textbf{THE MODELS AND THE BASIC EQUATIONS}}

We consider the model, which is characterized by the exotic
equation of state,

\begin{equation}
                \frac{p}{\rho} =  - w (r)
            \label{Eq1}
          \end{equation}

where $ w (r) $ is a positive function of radial coordinate.

A static spherically symmetric Lorentzian wormhole can be
described by a manifold $ R^2 X S^2 $ endowed with the general
metric in Schwarzschild co-ordinates $( t,r,\theta,\phi )$ as

\begin{equation}
                ds^2 = - e^{2f(r)} dt^2 + \frac{1}{[1 - \frac{b(r)}{r}]}dr^2+r^2 d\Omega_2^2
            \label{Eq1}
          \end{equation}

where,   $ r     \epsilon   (-\infty , +\infty) $ . \\

To describe a wormhole, the redshift function f (r) should be
finite and the shape function obeys the following properties
\begin{equation}
               b(r_0) = r_0
            \label{Eq1}
          \end{equation}
where $r_0$ is the throat of the wormhole.
\begin{equation}
               b^\prime(r_0) < 1
            \label{Eq1}
          \end{equation}
          \begin{equation}
               b(r) < r ,
                     r > r_0
            \label{Eq1}
          \end{equation}
Also the spacetime is asymptotically flat i.e. $\frac{b(r)}{r}
\rightarrow 0 $ as $ \mid r \mid \rightarrow \infty $.

 According
to Morris and Thorne [1], we assume $ f = constant $, to make the
problem simpler. This assumption implies that a traveller feels a
zero tidal force. This supposition would help for an advanced
engineer to construct a traversable passage.

Using the Einstein field equations
 $G_{\mu\nu} = 8\pi T_{\mu\nu} $, in orthonormal reference frame
 ( with $ c = G = 1 $ ) , we obtain the following stress energy
scenario,

\begin{equation}
                \rho(r) =\frac{b^\prime}{8\pi r^2}
                \label{Eq2}
          \end{equation}

\begin{equation}
                p(r) =\frac{1}{8\pi} \left[ - \frac{b}{r^3} \right]
                \label{Eq3}
          \end{equation}

\begin{equation}
                p_{tr}(r) =\frac{1}{8\pi} \left( 1 -
                \frac{b}{r} \right)  \left[ \frac{(- b^\prime r
              +
              b)
              }{2r^2(r-b)} \right]
                \label{Eq4}
          \end{equation}

where $\rho(r) $ is the energy density, $p(r)$ is the radial
pressure and $p_{tr}(r)$ is the transverse pressure. \\

Using the conservation of stress energy tensor $
T^{\mu\nu}_{;\nu} = 0 $, one can obtain the following equation

\begin{equation}
                p^\prime  + \frac{2}{r}p -
                \frac{2}{r}p_{tr} = 0
                \label{Eq5}
          \end{equation}

Now from equation (1), by using (6) and (7), one gets,

\begin{equation}
                \frac{p}{\rho} =  - w (r) =  -   \frac{b}{r b^\prime }
            \label{Eq1}
          \end{equation}

          \pagebreak

\cen{ \bf 3. \textbf{TOY MODELS OF WORMHOLES}}

Now, we will  discuss several toy models of wormholes:

\textbf{Specialization 1 : $ w (r) = w ( constant )$.}

Now consider the special case, $ w (r) = w ( constant )$, then
equation (10) yields,

\begin{equation}
                b =  b_0 r ^ { \frac{1}{w}}
            \label{Eq1}
          \end{equation}

[ $b_0$ is an integration constant ]

\begin{figure}[htbp]
    \centering
        \includegraphics[scale=.8]{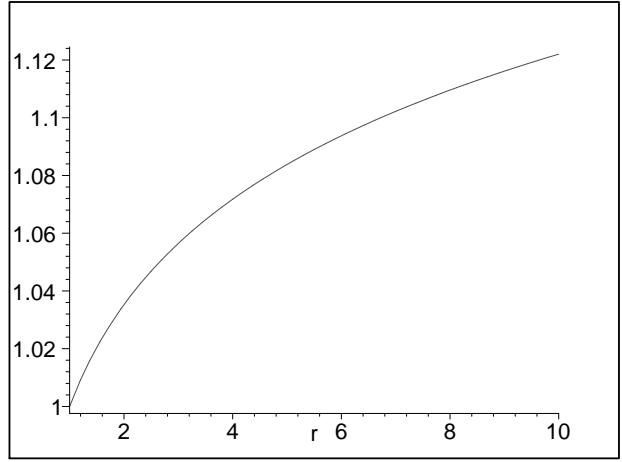}
        \caption{Diagram of the shape function of the wormhole }
   \label{fig:shape1}
\end{figure}

 Since the spacetime is asymptotically flat i.e.
$\frac{b(r)}{r} \rightarrow 0 $ as $ \mid r \mid \rightarrow
\infty $, then  the equation (11) is consistent only when $ w > 1
$.

The throat of the wormhole occurs at

\begin{equation}
                r =  r_0 = b_0 ^ { \frac{w}{w - 1}}
            \label{Eq1}
          \end{equation}

          Now we match the interior wormhole metric to the exterior
Schwarzschild metric . To match the interior to the exterior, we
impose the continuity of the metric coefficients, $ g_{\mu\nu} $,
across a surface, S , i.e. $ {g_{\mu\nu}}_{(int)}|_S =
{g_{\mu\nu}}_{(ext)}|_S $.

[ This condition is not sufficient to different space times.
However, for space times with a good deal of symmetry ( here,
spherical symmetry ), one can use directly the field equations to
match [10] ]

 The wormhole metric is continuous from the throat, $ r = r_0$
to a finite distance $ r = a $. Now we impose the continuity of $
g_{tt} $ and $ g_{rr}$,

$ {g_{tt}}_{(int)}|_S =  {g_{tt}}_{(ext)}|_S $

$ {g_{rr}}_{(int)}|_S =  {g_{rr}}_{(ext)}|_S $

at $ r= a $ [ i.e.  on the surface S ] since $ g_{\theta\theta} $
and $ g_{\phi\phi}$ are already continuous. The continuity of the
metric then gives generally

$ {e^{2f}}_{int}(a) = {e^{2f}}_{ext}(a) $ and $
{g_{rr}}_{({int})}(a) = {g_{rr}}_{({ext})}(a) $.

Hence one can find

\begin{equation}e^{2f}=  \left( 1 - \frac{2GM}{a} \right) \end{equation}

and $  1 - \frac{b(a)}{a} =  \left( 1 - \frac{2GM}{a} \right)  $
i.e. $ b(a) = 2GM $

This implies

$ b_0 a^{\frac{1}{w}} = 2GM $

Hence,
\begin{equation} a =  \left(\frac{2GM}{b_0} \right)^w \end{equation}
i.e. matching occurs at $ a =  \left(\frac{2GM}{b_0} \right)^w  $.

The interior metric $ r_0 < r \leq a $ is given by

\begin{equation}
               ds^2=  -  \left[ 1- b_0 a^{\frac{1-w}{w}} \right]dt^2+ \frac{dr^2}{ \left[ 1-b_0 r^{\frac{1-w}{w}} \right]}+r^2( d\theta^2+sin^2\theta
               d\phi^2)
          \end{equation}

The exterior metric $ a \leq r < \infty   $ is given by

\begin{equation}
               ds^2=  -  \left[ 1- \frac{b_0 a^{\frac{1}{w}}}{r} \right]dt^2+ \frac{dr^2}{ \left[ 1- \frac{b_0 a^{\frac{1}{w}}}{r} \right]}+r^2( d\theta^2+sin^2\theta
               d\phi^2)
          \end{equation}

\pagebreak

 Here, one can see that the null energy condition is
violated, $p + \rho < 0$ and consequently all the other energy
conditions. Now we will check whether the wormhole geometry is, in
principle, suffered by arbitrary small amount averaged null energy
condition (ANEC) violating phantom energy.  The ANEC violating
matter can be quantified by the integrals $ I = \oint  \rho  dV $,
$ I = \oint ( p_i + \rho ) dV $. In the model, we have assumed
that the ANEC violating matter is related only to $p$ ( radial
pressure ), not to the transverse components [ as one can see from
field equations (6) - (8), $p_{tr} =   \left(\frac{-1 + w(r) }{2}
\right)\rho$ and $w(r) > 1$ ].

According to Visser et al [11] , the information about the 'total
amount' of ANEC violating matter in the spacetime is given by the
integral,

\begin{equation}
                I = \oint ( p + \rho ) dV = 2\int_{r_0}^{\infty}( p + \rho
                )4\pi r^2 dr  \end{equation}

[ $ dV = r^2 \sin \theta dr d\theta d \phi $, factor two comes
from including both wormhole mouths ]

From the field equations, one can get,

\begin{equation}
                  p + \rho  = \frac{1}{8\pi r }
                    \left( 1- \frac{b}{r} \right )  \left[  \ln \frac{1} {  \left( 1-\frac{b}{r} \right)} \right]^\prime    \end{equation}

Hence,

\begin{equation} I=
                 \left[ ( r - b )\ln  \left(\frac{r}{r - b} \right) \right]_{r_0} ^{\infty}
                - \int_{r_0}^{\infty} \left[ ( 1 - b^\prime )\ln  \left(\frac{r}{r -
                b} \right) \right]dr
            \label{Eq1}
          \end{equation}

For the first expression, we see  that, at the throat $r_0$, $ b (
r_0) = r_0 $, the boundary term at $r_0$ vanishes. Now we consider
the boundary term at infinity.

Let us denote as $\chi$ the contribution into this term from
infinity. Then,

  \begin{equation}
                \chi  = \lim_{r\rightarrow \infty } (
                r-b)\ln\left[\frac{r}{r-b}\right]
            \label{Eq1}
          \end{equation}
This can be rewritten as
\begin{equation}
                \chi  = \lim_{r\rightarrow \infty } r \left(
                1-\frac{b}{r} \right)\ln \left[\frac{r}{r-b}
                \right]
            \label{Eq1}
          \end{equation}
 Now, as in this limit the quantity$ \frac{b}{r}$ is small, we may expand the logarithm as
 \linebreak
 $\ln \left( 1 - \frac{b}{r} \right) = - \frac{b}{r}+ ...$, where only the main
 term is retained. Neglecting here the term   $ \frac{b}{r}$ in
 parentheses, one obtains $ \chi = b(\infty)$.
 Here, $  b(\infty) = \infty $.   Hence, in this case the total amount of ANEC violating matter
 is infinitely large. This case is not physically interesting.

\textbf{Specialization 2: $ w(r) = A r^n$ }

In this case, we consider $ w(r) = A r^n $ , where n and $A$ are
two positive constants. For this consideration, equation (10)
gives,

\begin{equation}
                b  = exp\left(B-\frac{1}{A n r^n}\right)
            \label{Eq1}
          \end{equation}
where B is an integration constant.

We assume, throat of the wormhole occurs at $r=r_0$, then
$b(r_0)=r_0$ implies

\begin{equation}
                B  = \frac{1}{A n r_0^n } + \ln r_0
            \label{Eq1}
          \end{equation}

So, the shape function  takes the following form as

\begin{equation}
                b  = exp\left( \frac{1}{A n r_0^n  } + \ln r_0-\frac{1}{n A r^n}\right)
            \label{Eq1}
          \end{equation}

Since, $n>0$, $w>1$,  for all $r>r_0>[\frac{1}{A}]^{\frac{1}{n}}$.
So the assumption $ w(r) = A r^n$ is justified to explore the
phantom energy with 'r' dependent equation of state. Here the
space time is asymptotically flat i.e. $\frac{b(r)}{r}\rightarrow
0 $ as $ \mid r \mid \rightarrow \infty $.

\begin{figure}[htbp]
    \centering
        \includegraphics[scale=.4]{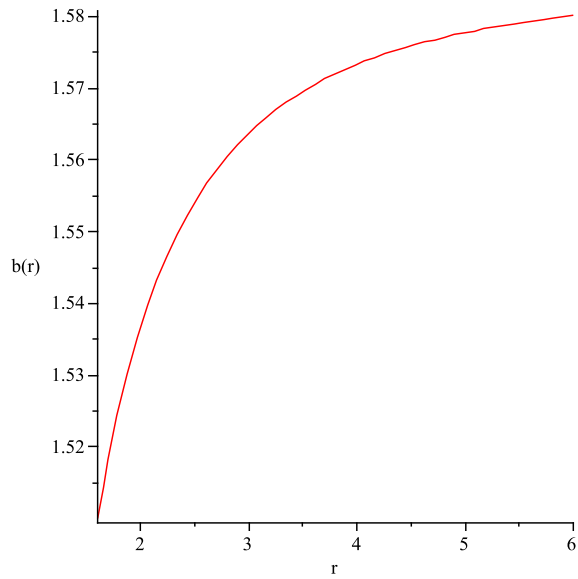}
        \caption{Diagram of the shape function of the wormhole}
   \label{fig:shape2}
\end{figure}

\begin{figure}[htbp]
    \centering
        \includegraphics[scale=.4]{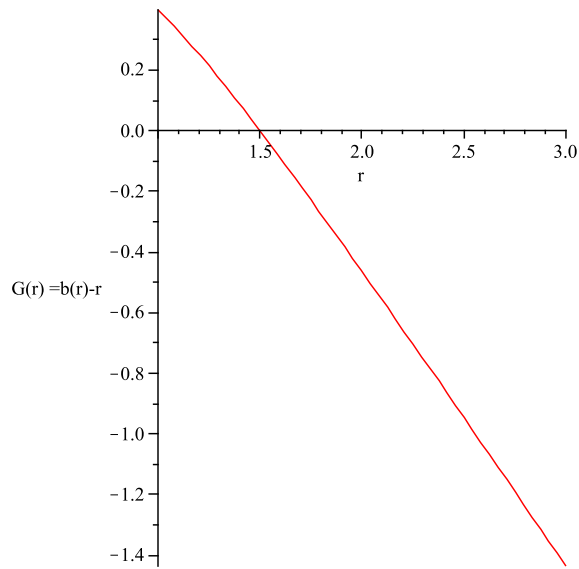}
        \caption{Throat occurs where $G(r)$ cuts 'r' axis i.e. at
        $ r = 1.5 $
        ( choosing suitably the parameters as $r_0 = 1.5$, $A=4$
    and  $n = 2 $ ).}
   \label{fig:shape2}
\end{figure}

From the graph ( fig. 3 ), one can also note that  when $r>r_0 $,
$G(r)< 0$ i.e. $  b(r) -r < 0 $. This implies $ \frac{b(r) }{r} <
1 $ when $r>r_0 $. Also,  from the graph, we see that G is a
decreasing function of r for $r \geq r_0$ and hence
$G^{\prime}(r)< 0$ for $r \geq r_0$. In other words,
$b^{\prime}(r_0)< 1$  i.e. flare-out condition has been satisfied.
Thus obtained shape function would represent a wormhole structure.

Now we can match this interior wormhole metric with exterior
Schwarzschild metric at $a$  where $ a = \left[ \frac{1}{An\left(
\frac{1}{Anr_0^n} + \ln { \frac{r_0}{2GM}}\right) }
\right]^\frac{1}{n} $

Here the interior metric $ r_0 < r \leq a $ is given by
\begin{equation}
               ds^2=  - \left[ 1- \frac{exp\left( \frac{1}{Anr_0^n } + \ln r_0-\frac{1}{n Aa^n}\right)}{a} \right]dt^2+
               \frac{dr^2}{\left[ 1- \frac{exp\left( \frac{1}{Anr_0^n } + \ln r_0-\frac{1}{n Aa^n}\right)}{r} \right]}+r^2(
d\theta^2+sin^2\theta
               d\phi^2)
          \end{equation}

The exterior metric $ a \leq r < \infty   $ is given by
\begin{equation}
               ds^2=  - \left[ 1- \frac{exp\left( \frac{1}{Anr_0^n } + \ln r_0-\frac{1}{n Aa^n}\right)}{r} \right]dt^2+
               \frac{dr^2}{\left[ 1- \frac{exp\left( \frac{1}{Anr_0^n } + \ln r_0-\frac{1}{n Aa^n}\right)}{r} \right]}+r^2(
d\theta^2+sin^2\theta
               d\phi^2)
          \end{equation}

In this case, we are interested to measure the total amount of
ANEC violating matter. We consider the wormhole field deviates
from the throat out to a radius $a$. Thus the total amount of ANEC
violating matter is to matching the interior solution to an
exterior spacetime at $a$. Then the volume integral takes the
value,

$
                I =  [  b(a) -a]\left[   \ln \left( 1 - \frac{b(a)}{a}\right)\right] + (a -r_0)-
    - (a\ln a - r_0 \ln r_0)-
[  b(a) -a]\left[   \ln \left(  \frac{a-b(a)}{e}\right)\right]+
[b(a)\ln a - b(r_0) \ln r_0]-\ln\frac{a}{r_0}
               -\frac{1}{An^2} \left( \frac{1}{a^n} -
               \frac{1}{r_0^n}\right) + .............
         \label{Eq1}$

          \begin{equation}   \end{equation}

  This implies that  the total amount of ANEC
violating matter depends on several parameters, namely, $a, n, A,
r_0 $. If we kept fixed the parameters $n, r_0, A $, then the
parameter $a$ plays significant role to reducing total amount of
ANEC violating matter. Thus total amount of ANEC violating matter
can be made small by taking suitable position, where interior
wormhole metric will match with exterior Schwarzschild metric.
This proves that it is possible to construct wormhole with
 small amount of phantom energy characterized by
variable equation of state parameter.

\textbf{Specialization 3 : Specific shape function : $ b (r) = D
 \left(1 - \frac {A}{r}  \right) \left(1 - \frac {B}{r}  \right)$.}

Consider the specific form of the shape function as

\begin{equation}
                b (r) = D  \left(1 -
\frac {A}{r}  \right) \left(1 - \frac {B}{r}  \right)
            \label{Eq1}
          \end{equation}

where A, B and D($ > 0 $) are arbitrary constants.

\begin{figure}[htbp]
    \centering
        \includegraphics[scale=.8]{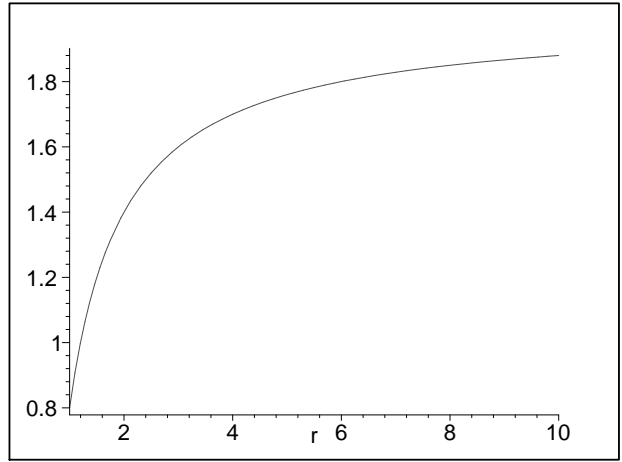}
        \caption{Diagram of the shape function of the wormhole}
   \label{fig:shape2}
\end{figure}

\pagebreak

 For this case, the equation of state parameter
function takes the form

\begin{equation}
                w(r) = \frac {[(r-A)(r - B)]}{[( A + B)r - 2AB]}
            \label{Eq1}
          \end{equation}

\begin{figure}[htbp]
    \centering
        \includegraphics[scale=.8]{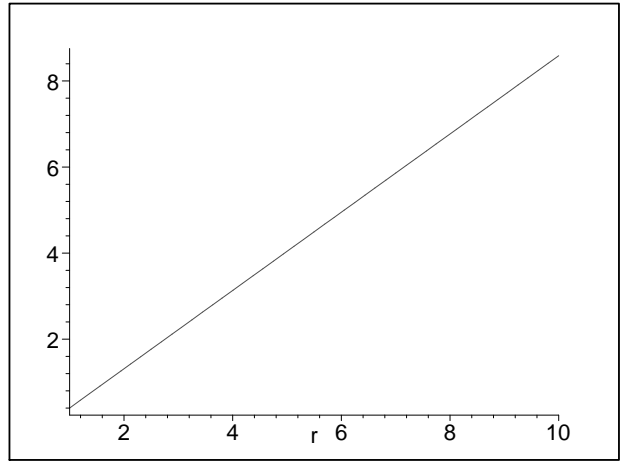}
        \caption{Diagram of the Equation of state parameter. Here r can not be taken arbitrarily large. The fig. is limited by values  $ \leq a $,
         where interior wormhole metric will match with
  exterior Schwarzschild metric. }
   \label{fig:shape3}
\end{figure}

 We will now verify whether the particular choice of the
shape function would represent the wormhole structure. One can see
easily that $\frac{b(r)}{r} \rightarrow 0 $ as $ \mid r \mid
\rightarrow \infty $.

Throat of the wormhole occurs at $ r = r_0 $ , where $r_0$
satisfies the following equation $ b(r_0) = r_0$ i.e. $r_0^3 -
Dr_0^2 + ( A + B) D r_0 - ABD =0$.

 The solution of this
equation is
\begin{equation}
                r_0 = S + T + \frac {D}{3}
            \label{Eq1}
          \end{equation}
where ,

 $ S = [ R + \sqrt{Q^3 + R^2}]^{\frac{1}{3}} $ and $  T = [ R - \sqrt{Q^3 +
 R^2}]^{\frac{1}{3}}$,

 $ Q = \frac {3D(A + B) - D^2}{9},  R = \frac {27ABD + 2D-9D^2(A +
 B)}{54}$

 Since $r_0$ is a root of the above equation
, then by standard theorem of algebra, either $g(r)\equiv b(r) -
r  < 0$ for $ r
> r_0$ and $g(r) > 0 $ for $ r < r_0 $ or
$g(r) > 0$ for $ r > r_0 $ and $g(r) < 0$ for $ r < r_0 $. Let us
take the first possibility and one can note that for $ r > r_0 $
, $g(r) < 0$, in other words, $b(r) < r $. But when $r<r_0$, $g(r)
> 0$, this means, $b(r)
> r $, which violates the wormhole structure given in equation(2).

Now we are  matching our interior wormhole metric with the
exterior Schwarzschild metric at $a$ where

 $ a = \frac{AD + BD +
\sqrt{( AD + BD )^2 - 4ABD( D - 2GM ) }}{2(D - 2GM)}$.

Here the interior metric $ r_0 < r \leq a $ is given by
\begin{equation}
               ds^2=  - \left[ 1- \frac{D}{a} \left(1 -
\frac {A}{a} \right)\left(1 - \frac {B}{a} \right)\right]dt^2+
\frac{dr^2}{\left[ 1- \frac{D}{r} \left(1 - \frac {A}{r}
\right)\left(1 - \frac {B}{r}\right )\right]}+r^2(
d\theta^2+sin^2\theta
               d\phi^2)
          \end{equation}

The exterior metric $ a \leq r < \infty   $ is given by
\begin{equation}
               ds^2=  - \left[ 1- \frac{D}{r} \left(1 -
\frac {A}{a} \right)\left(1 - \frac {B}{a} \right)\right]dt^2+
\frac{dr^2}{\left[ 1- \frac{D}{r} \left(1 - \frac {A}{a}
\right)\left(1 - \frac {B}{a} \right)\right]}+r^2(
d\theta^2+sin^2\theta
               d\phi^2)
          \end{equation}

In this case, the total amount of ANEC violating matter in
spacetime with a cutoff of the stress energy at $a$ is given by
 \begin{equation}
                I = D + [  a - b(a) ]\left[ 1 + \ln \left( 1 - \frac{b(a)}{a}\right)\right] + ( a - r_0) -
                D\ln \frac{a}{r_0} - D ( A +B )\left( \frac{1}{a} -\frac{1}{r_0}\right)-
                   2ABD\left(\frac{1}{a^2}-\frac{1}{r_0^2}\right)
         \label{Eq1}
          \end{equation}
  This implies that  the total amount of ANEC
violating matter depends on several parameters, namely, $a, A, B,
D, r_0 $. If we kept fixed the parameters $A, B, D, r_0 $, then
the parameter 'a' plays significant role to reducing total amount
of ANEC violating matter. Thus total amount of ANEC violating
matter can be made small by taking suitable position, where
interior wormhole metric will match with exterior Schwarzschild
metric. This proves that it is possible to construct wormhole with
 small amount of phantom energy characterized by
variable equation of state parameter.

 According to Morris and Thorne [1] , the 'r'
co-ordinate is
ill-behaved near the throat, but proper radial distance\\
\begin{equation}
 l(r) = \pm \int_{r_0^+}^r \frac{dr}{\sqrt{1-\frac{b(r)}{r}}}
            \label{Eq20}
          \end{equation}
 must be well behaved everywhere i.e. we must require that $ l(r)
 $is finite throughout the space-time . \\
 For our model, ( taking $B = 0$ ), one can determine the proper
 distance through the wormhole as
\begin{equation}
 l(r) = \sqrt{r^2 - Dr + DA}  + \frac{D}{2} \ln \left[ \frac{2\sqrt{r^2 - Dr +
 DA} + 2r - D }{ 2r_0 - D }\right]
             \label{Eq20}
          \end{equation}
\begin{figure}[htbp]
    \centering
        \includegraphics[scale=.8]{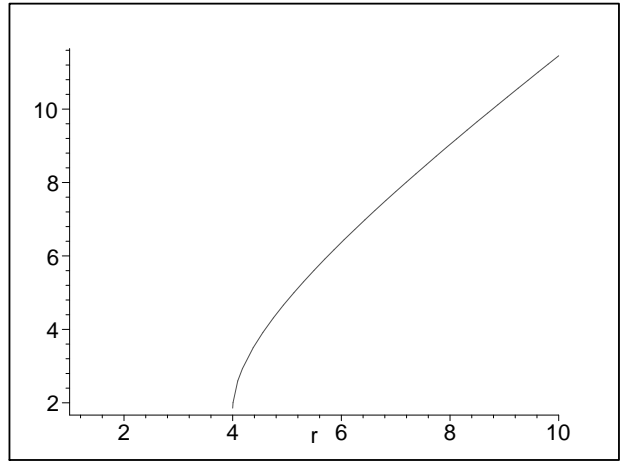}
        \caption{Diagram of the radial proper distance ( D = 2, A = - 4, $r_0 = 4$ )}
   \label{fig:shape4}
\end{figure}
 The radial proper distance is measured from $r_0$ to
any $ r > r_0$. Note that on the throat $r = r_0$,  $ l= 0$.

\textbf{Specialization 4 : Specific shape function : $ b (r) = A
\tanh Cr $.}

Now we make the specific choice for the shape function as

\begin{equation}
                b (r) =A \tanh Cr
            \label{Eq1}
          \end{equation}

where A ($ > 0 $) and C are arbitrary constants.

\begin{figure}[htbp]
    \centering
        \includegraphics[scale=.8]{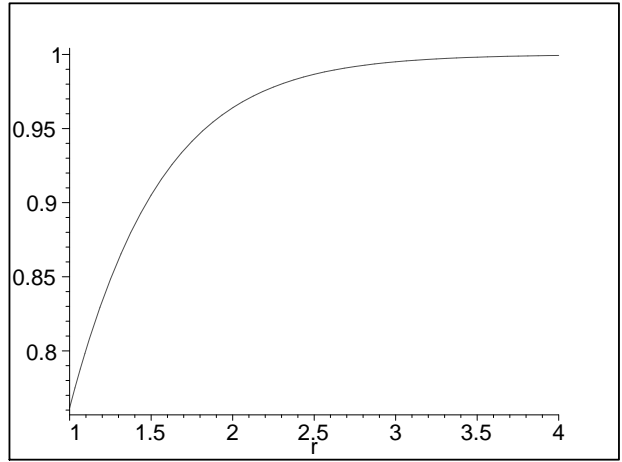}
        \caption{Diagram of the shape function of the wormhole}
   \label{fig:wh20}
\end{figure}

Using the equation (10), one gets
\begin{equation}
                w(r) =\frac{C}{2r} \sinh 2Cr
            \label{Eq1}
          \end{equation}
\begin{figure}[htbp]
    \centering
        \includegraphics[scale=.8]{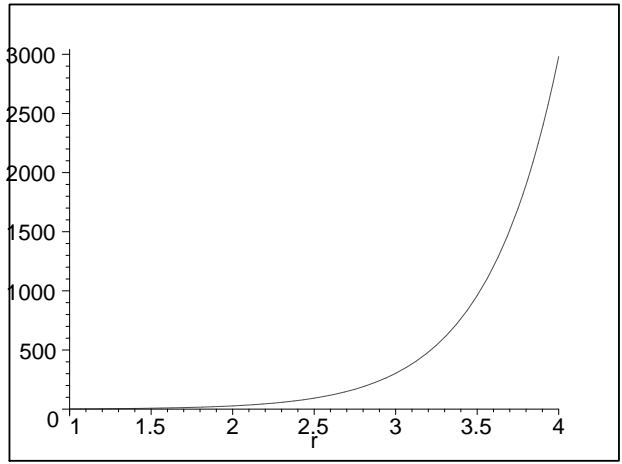}
        \caption{Diagram of the Equation of state parameter. Here r can not be taken arbitrarily large. The fig. is limited by values  $ \leq a $,
         where interior wormhole metric will match with
  exterior Schwarzschild metric.}
   \label{fig:shape3}
\end{figure}
\\
\\
\\
It is easy to verify the above particular choice of the shape
function would represent wormhole structure. Here, $\frac{b(r)}{r}
\rightarrow 0 $ as $ \mid r \mid \rightarrow \infty $ and throat
occurs at $ r= r_0 $ for which $  b(r_0) = r_0$ i.e. $A\tanh Cr_0
= r_0$.
\\
\\

[ If one chooses $ A = 2$ and $C=1$, the graph of the function
$F(r)=  b(r) -r $ indicates the point $r_0$ where $F(r)$ cuts the
'r' axis. From the graph, one can also note that when $r>r_0 $,
$F(r)< 0$ i.e. $  b(r) -r < 0 $. This implies $ \frac{b(r) }{r} <
1 $ when $r>r_0 $.]
\\
\\

\begin{figure}[htbp]
    \centering
        \includegraphics[scale=.9]{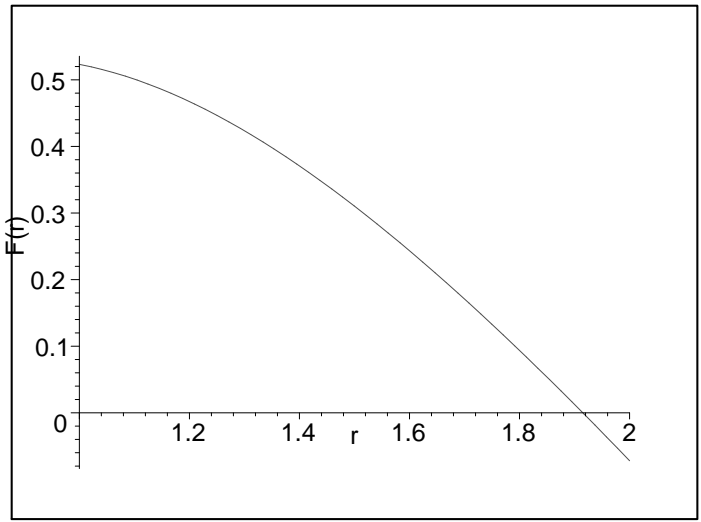}
        \caption{Throat occurs where $F(r)$ cuts 'r' axis}
   \label{fig:wh20}
\end{figure}

\pagebreak

 Now matching this interior metric of wormhole with the
exterior Schwarzschild metric at a, where $ a = \frac{1}{2C} \ln
\frac{A + 2GM}{A - 2GM} $, one gets, the interior metric $ r_0 < r
\leq a $ as

\begin{equation}
               ds^2=  - [ 1- \frac{A\tanh Ca}{a} ]dt^2+ \frac{dr^2}
               {[1- \frac{A\tanh Cr}{r}]} + r^2( d\theta^2 + sin^2\theta
               d\phi^2)
          \end{equation}

Here the exterior metric $ a \leq r < \infty   $ is given by

\begin{equation}
              ds^2=  - [ 1- \frac{A\tanh Ca}{r} ]dt^2+ \frac{dr^2}
               {[1- \frac{A\tanh Ca}{r}]} + r^2( d\theta^2 + sin^2\theta
               d\phi^2)
          \end{equation}

In this case, the amount of ANEC violating matter in the spacetime
with a cut off of the stress energy at 'a' is given by

$
                I = A + [  a - A\tanh Ca ]\ln {a} + ( a - r_0) -
                A[ C(a - r_0) - \frac{C^3}{9} ( a^3 - r_0^3) +
\frac{2C^5}{75} ( a^5 - r_0^5)- ..... \linebreak
\frac{(-1)^{n-1}B_n 2^{2n} ( 2^{2n} - 1)}{(2n - 1)(2n!)}C^{2n-1} (
a^{2n-1} - r_0^{2n-1})+ ....] + (  a - A\tanh Ca ) \ln[(a - A\tanh
Ca) - 1]
       $

    \begin{equation}   \end{equation}

where, $B_n$ is the Bernoulli number.

Also, in this case, if we treat the parameters $A, C, r_0 $ as
fixed constants, then total amount of ANEC violating matter can be
reduced  to small quantity by taking suitable position, where
interior wormhole metric will match with exterior Schwarzschild
metric. In other words, this type of wormhole can be constructed
with
 small quantity of ANEC violating phantom energy
material.

\pagebreak

\cen{ \bf 4. \textbf{CONCLUDING REMARKS}}

  Our
aim in this paper is  to provide a prescription for obtaining
wormhole where stress energy tensor is characterized by phantom
energy with variable equation of state parameter. We have provided
several toy models according to this new proposal. In the first
two models we have considered phantom energy sources that give
birth wormhole like structure whereas last two models, we have
considered specific forms of the shape functions of the wormhole
and try to search matter sources ( phantom like ) that generate
the
above wormhole structures.\\
As mentioned above, to be a wormhole solution, the condition
$b^\prime(r_0) < 1 $ is to be imposed. Now for the first case,
$b^\prime(r_0) = \frac {1}{w} < 1 $, since $ w > 1 $,
corresponding to solution(11). For the second case,  $
b^\prime(r_0) < 1 $
 implies $r_0>(\frac{1}{A})^{\frac{1}{n}}$ , corresponding to
solution(22) and for the last cases, one has to assume $r_0
> (A + B) \pm \sqrt{(A + B)^2 - 3AB}$ and $ r_0 > \frac{1}{C} \cosh^{-1}
\sqrt{AC}$, corresponding to the solutions (28) and (30)
respectively.\\ We have established a matching each of four
interior wormhole metrics with the exterior Schwarzschild
metric.\\
Except model 1, all the other models reveal the fact that one may
construct wormholes with small amounts of phantom energy as
desired which is characterized by variable equation of state
parameter.\\ The effective mass inside the radius 'r' is defined
by $M(r)= \frac {b(r)}{2}$ and the limit,  $ \lim _ {r
\rightarrow\infty } M(r)= M $, if exists, represents the
asymptotic wormhole mass seen by an distant observer. In the first
case, this limit does not exist whereas for the last three cases,
one can see that M exists and equal to $ \frac{1}{2} exp\left(
\frac{1}{Anr_0^n } +
               \ln r_0\right)$
 for the second case and equal to $\frac{D}{2}$ and $\frac{A}{2}$ for the last cases. This
implies that a distant observer could not see any difference of
gravitational nature between Wormhole and a compact mass 'M'. \\
The assumption that the redshift function to be constant function
implies the tidal gravitational force experienced by a traveller
is zero. Thus one of the traversibility condition is satisfied in
other words, our wormholes are traversable. Hence our wormholes
containing small amount of exotic matter in spite of they are
traversable for human beings.\\
The specializations 2-4 are taken from intuition and simplicity
from the mathematical point of view. Since the proposed phantom
energy is characterized by variable EOS parameter, so there are
thousands  number of choices may be considered. Before selecting
the specializations 2-4, we are dealing with several choices but
we find these are the physically acceptable wormhole models ( i.e.
satisfying all the characteristics of wormhole as well as
supported by small amount of ANEC violating matter ). We hope
scientists would be motivated by our approach and in future, will
try to find sophisticated  way for constructing wormholes.

{ \bf Acknowledgements }

          F.R. is thankful to Jadavpur University and DST , Government of India for providing
          financial support. MK has been partially supported by
          UGC,
          Government of India under MRP scheme Finally, we are thankful to the anonymous referee for his
critical remarks and constructive suggestions. .

\pagebreak


\end{document}